\newtheorem{theorem}{Theorem}
\newtheorem{corollary}{Corollary}[theorem]
\newtheorem{lemma}{Lemma}[theorem]
\documentclass[11pt]{article}
\usepackage{graphicx}
\usepackage{amsfonts}
\begin{document}
\thispagestyle{empty}
\setcounter{page}{1}
\setlength{\baselineskip}{1.5\baselineskip}
\title{Efficient On-line Schemes for Encoding Individual Sequences with Side Information at the Decoder \thanks{This research is supported by the Israeli Science Foundation (ISF),
grant no.\ 208/08.}}
\author{Avraham Reani and Neri Merhav\\
Department of Electrical Engineering\\
Technion - Israel Institute of Technology\\
Technion City, Haifa 32000, Israel\\
Emails: [avire@tx,merhav@ee].technion.ac.il}
\maketitle
\begin{abstract}
We present adaptive on-line schemes for lossy encoding of individual sequences under the conditions of the Wyner-Ziv (WZ) problem, i.e., the decoder has access to side information whose statistical dependency on the source is known. Both the source sequence and the side information consist of symbols taking on values in a finite alphabet $\cal{X}$. 
In the first part of this article, a set of fixed-rate scalar source codes with zero delay is presented. We propose a randomized on-line coding scheme, which achieves asymptotically (and with high probability), the performance of the best source code in the set, uniformly over all source sequences. The scheme uses the same rate and has zero delay. We then present an efficient algorithm for implementing our on-line coding scheme in the case of a relatively small set of encoders. We also present an efficient algorithm for the case of a larger set of encoders with a structure, using the method of the weighted graph and the Weight Pushing Algorithm (WPA). 
In the second part of this article, we extend our results to the case of variable-rate coding. A set of variable-rate scalar source codes is presented. We generalize the randomized on-line coding scheme, to our case. This time, the performance is measured by the Lagrangian Cost (LC), which is defined as a weighted sum of the distortion and the length of the encoded sequence. We present an efficient algorithm for implementing our on-line variable-rate coding scheme in the case of a relatively small set of encoders. We then consider the special case of lossless variable-rate coding. An on-line scheme which use Huffman codes is presented. We show that this scheme can be implemented efficiently using the same graphic methods from the first part.
Combining the results from former sections, we build a generalized efficient algorithm for structured set of variable-rate encoders.
Finally, we show how to generalize all the results to general distortion measures.
The complexity of all the algorithms is no more than linear in the sequence length.
\end {abstract}
\clearpage
\section{Introduction}
Consider a communication system with the following components: an individual source sequence to be compressed, a discrete memoryless channel (DMC) with known statistics, a noiseless channel with rate constraint $R$, and a decoder. The encoder maps the source sequence, $x_1,x_2,\ldots,x_n$, into a sequence of channel symbols, $z_1,z_2,\ldots,z_n$, taking values in $\{1,2,\ldots,M\}$, $M=2^R$, which is transmitted to the decoder via a noiseless channel. The decoder, in addition to the encoded data arriving from the noiseless channel, has access to a side information sequence, $y_1,y_2,\ldots,y_n$, which is the output of the DMC fed by the source sequence. Using the compressed data, $z_1,z_2,\ldots,z_n$, and the side information, the decoder produces a reconstructed sequence $\hat{x}_1,\hat{x}_2,\ldots,\hat{x}_n$. The goal is to minimize the distortion between the source and the reconstructed signal by optimally designing an encoder-decoder pair. This is a variation of the problem of rate-distortion coding with decoder side information, which is well known as the Wyner-Ziv (WZ) coding problem, first introduced in [6]. The case of scalar source codes for the WZ problem was handled in several papers, e.g. [7] and [8]. In contrast to our case, these schemes operate under specific assumptions of known source statistics. WZ coding of individual sequences was also considered, e.g. in [9] and [10], and existence of universal schemes was established. However, these schemes are based on block coding or DUDE implementation and assume the knowledge of the source and side information sequences in advance. Thus, they are irrelevant to the case of on-line encoding considered here.

A coding scheme is said to have an overall delay of no more than $d$ if there exist positive integers $d_1$ and $d_2$, with $d_1+d_2\leq d$, such that each channel symbol at time $t$, $z_t$, depends only on $x_1,\ldots,x_{t+d_1}$, and each reconstructed symbol $\hat{x}_t$ depends only on $z_1,\ldots,z_{t+d_2}$ and $y_1,\ldots,y_{t+d_2}$.
Weissman and Merhav [3], following Linder and Lugosi [2], constructed a randomized limited delay lossy coding scheme for individual sequences using methods based on prediction theory. These schemes perform, for any given reference class of source codes, called $experts$, almost as well as the best source code in the set, for all individual sequences. The performance of the scheme is measured by the distortion redundancy, defined as the difference between the normalized cumulative distortion of the scheme and that of the best source code in the set, matched to the source sequence. The scheme is based on random choices of source codes from the set. The random choices are done according to exponential weights assigned to each code. The weight of each source code, each time we choose a code, depends on its past performance and has to be calculated. Thus, implementing this scheme for a large set of source codes requires efficient methods, to prevent prohibitive complexity. Gy\"{o}rgy, Linder and Lugosi offered efficient algorithms for implementing such a scheme for sets of scalar quantizers [4],[5] without side information. Our main contribution in this paper is to extend this scenario to include side information at the decoder, in the spirit of the WZ problem, for both the fixed rate case and the variable-rate case.

In the first part of this paper, a fixed-rate, zero-delay adaptive coding scheme for individual sequences under the WZ conditions is presented.
We define a set of scalar source codes for the WZ problem. Then, the scheme of [3] is extended for the WZ problem w.r.t. the Hamming distortion measure. For any given set of WZ source codes, this scheme performs asymptotically as well as the best source code in the set, for all source sequences. We then demonstrate efficient implementations of this scheme. First, it is shown that the scheme can be implemented efficiently for any relatively small set of encoders, even though the set of decoders is large. Then, using graph-theoretic methods similarly to [5], we show that we can implement the scheme for large sets of scalar encoders with a structure.

In the second part of the paper, we extend the results of [3], and the coding schemes from the first part, to the variable-rate coding case.
Without loss of generality, we assume that the noiseless channel is binary. The encoder, instead of using a fixed-rate code for encoding the source sequence into $M$ symbols, now uses a variable-length binary prefix code with $M$ codewords. The decoder, upon receiving the binary encoded sequence, first produces the indexes of the codewords transmitted, and then continues exactly as in the fixed-rate case. The prefix property enables instantaneous  decoding of the codewords.

We start by defining a set of variable-rate scalar source codes. Then, the scheme of [3] is generalized to the variable-rate case. The performance is now measured by the LC function, which is defined as a weighted sum of the distortion and the length of the binary encoded sequence.
As before, for any given set of variable-rate source codes, this scheme performs asymptotically as well as the best source code in the set, for all source sequences. We then demonstrate efficient implementations of this scheme. Again, it is shown that the scheme can be implemented efficiently for any relatively small set of encoders, in a way similar to the fixed-rate case. Then, we handle the special case of lossless variable-rate coding. We first demonstrate a method of representing sets of Huffman codes on an a-cyclic directed graph. Using this representation and the WPA, we present efficient implementation for the lossless case. Then, combining this result and the set of encoders with a structure from the fixed-rate case, we show that we can implement the generalized variable-rate scheme for large sets of scalar encoders. Finally, all the implementations are generalized to accommodate any distortion measure, at the price of increased complexity.

It should be pointed out that the development of our efficient on-line scheme in the fixed-rate case, is not a straightforward extension of those in [4],[5] because of the following reasons: (i) Due to the side information, the optimal partition of the source alphabet does not necessarily correspond to intervals. (ii) The problem of choosing an expert is more complicated in the WZ setting. In [4] and [5], which deals with quantizers, the problem of choosing an expert is reduced to the problem of choosing decoding points. Given these points, the encoder is chosen to be the nearest-neighbor encoder which uses this points. In our case, this mechanism is irrelevant, and we have to choose the encoder and decoder separately.  

The remainder of the paper is organized as follows. In Section 2, a formal description of the fixed-rate case is given. In Subsection 2.1, we define the set of WZ scalar source codes. A general coding scheme, which achieves essentially the same performance as the best in a given set of WZ codes, is presented in Subsection 2.2. Section 3 is dedicated to the efficient implementation of this scheme for sets of scalar source codes. In Subsection 3.1, we present an efficient implementation for large sets of encoders with structure, using graphical methods. In Section 4, we give a formal description of the problem for the variable-rate case. In Subsection 4.1 we define the set of variable-rate source codes. In Subsection 4.2, we generalize the scheme and results of section 2. In Section 5, we present an efficient implementation of this scheme for scalar source codes with variable-rate coding. In Subsection 5.1 we handle the special case of lossless variable-rate coding. We establish efficient scheme which achieves essentially the same compression of the best in a given set of Huffman codes. In Subsection 5.2, we present an efficient implementation for the general lossy coding scheme. In Subsection 5.3, we show how to generalize our results to any bounded distortion measure. Finally, in Subsection 5.4, we describe the implementation of our variable-rate coding scheme, for the special case of quantizers which use Huffman codes.

\section{Definition of an on-line Adaptive WZ Scheme}
Throughout this paper, for any positive integer $n$, we let $a^n$ denote the sequence $a_1, a_2,\ldots,a_n$.

Given a source sequence $x^n$, the encoder maps the source sequence $x^n$ into a sequence $z^n$ whose symbols $\{z_i\}$ take on values in the set $\{1,2,\ldots,M\}$. The decoder, in addition to $z^n$, has access to a sequence $y^n$, dependent on $x^n$ via a known DMC, defined by the single-letter transition probability $P_{Y|X}(y_i|x_i)$, which is the probability of $y_i$ given $x_i$. Based on $z^n$ and $y^n$, the decoder produces the reconstructed sequence $\hat{x}^n$.
For convenience, we assume that $x_i$, $y_i$ and $\hat{x}_i$, all take on values in the same finite alphabet $\cal{X}$ with cardinality $|\cal{X}|$. All the results can be generalized straightforwardly to the case of different alphabets.
The distortion between two symbols is defined to be the Hamming distortion:
\begin{equation}
\rho(x,\hat{x}) = \left\{ \begin{array}{ll}
         0 & \mbox{if $x=\hat{x}$}\\
         1 & \mbox{elsewhere}.\end{array} \right.
\end{equation}   
We define the distortion for the input symbol $x_t$ at time $t$, $(t=0,1,2,\ldots)$ as:
\begin{equation}
\Delta_t(x_t)=\mathbb{E}\rho(x_t,\hat{x}_t(z^t,y^t))
\label{input_symbol_dis_def}
\end{equation}
where the expectation is taken with respect to $y^t$.
\subsection{Definition of the reference set of source codes}
In this part, we define a general set of scalar source codes, here referred to as {\it{experts}}. 
Each expert is a source code with a fixed rate, $R=\log M$, which partitions $\cal{X}$ into $M$ disjoint subsets $( m_1, m_2,\ldots, m_M)$. The encoder $e$ for each expert is given by a function $e : \cal{X} \rightarrow$$\{1, 2, \ldots, M\}$ that is, $z_i=e(x_i)$.
The decoder $d$ receives $z_i$, together with the side information $y_i$, and generates $\hat{x}_i$, using a decoding function 
$d : \{1, 2, \ldots,M\}\times\cal{X}\rightarrow\cal{X}$, i.e., $\hat{x}_i=d(z_i,y_i)$.
The definition above is not complete. It is easy to see that different encoders may actually implement the same partition.
For example: if $\cal{X}=$$\{1,2,3\}$ and $M=2$, consider the two encoders:
\begin{equation}
\begin{array}{llll}
e_1:&e_1(1)=1,&e_1(2)=2,&e_1(3)=2\\ 
e_2:&e_2(1)=2,&e_2(2)=1,&e_2(3)=1
\end{array}
\end{equation}
It is easy to see that they have the same functionality. In our definition we treat these encoders as the same encoder, otherwise, the same expert will be taken into account several times. The number of times depends on the specific partition, so we will get an unbalanced weighting of experts.
\subsubsection{Definition of the encoders using the partition matrix}
To define an encoder uniquely, and to get bounds on the cardinality of the general set of encoders, let us define the partition matrix:
\begin{equation}
PM_{j,l} = \left\{ \begin{array}{ll}
         1 & \mbox{if $x_j$ and $x_l$ belongs to the same subset.}\\
         0 & \mbox{else}.\end{array} \right.
\end{equation}        
where $j,l \in \{1,2,\ldots,|\cal{X}|\}$, are the indexes of the alphabet letters, $x_j,x_l\in {\cal{X}}$, given that we ordered the alphabet in some arbitrary order.

$\mathbf{The\ properties\ of\ PM}$:
\begin{enumerate}
	\item $PM_{i,j}\in(0,1)$.
	\item If $i=j$ then $PM_{i,j} = 1$.
	\item $PM$ is symmetric, i.e. $PM_{i,j} = PM_{j,i}$
	\item If $PM_{i,j} = 1$ and $PM_{i,k} = 1$ then $PM_{j,k} = 1$.
	\item If $PM_{i,j} = 1$ and $PM_{i,k} = 0$ then $PM_{j,k} = 0$.
\end{enumerate}
It is easy to see that each partition matrix (a matrix which has the above properties) defines unique partition of the alphabet thus defines an encoder uniquely.
Using the properties of this matrix, we can derive bounds on the number of encoders:
\begin{equation}  
2^{|{\cal{X}}|-1}\leq Number\ of\ PM's \leq 2^{|{{\cal{X}}}|\log{M}}
\end{equation}
The lower bound is derived from the fact that the first row to be determined has ${|\cal{X}|}-1$ degrees of freedom, i.e., it can be any binary vector of length $|\cal{X}|$ and with the first element equals to $1$. This reflects the fact, that the choice of the first subset of letters is unrestricted.
The upper bound is derived from the fact that the number of encoders without the limitation of counting every partition only once is $M^{|\cal{X}|}$.  
So the number of encoders is exponential in $|{\cal{X}}|$. Therefore, using the general set of encoders is a challenge from a computational complexity point of view.
\subsubsection{Definition of the decoders}
We limit our discussion to decoders which satisfy:
\begin{equation} 
d(e(\hat{x}),y)= \hat{x},\ for\ all\ \hat{x}\ and\ y.
\label{possible_decoders}
\end{equation}
which means that the decoded symbol $\hat{x}$, is chosen from the same subset of the input symbol $x$.
Using the Hamming distortion measure, it is easy to see that there is no point to choose $\hat{x}_i$ outside the subset $m_{z_i}$, hence this set of decoders is sufficient. For other distortion measures, the results can be generalized straightforwardly to the set of all possible decoders.
From the above definition, we see that every encoder defines a set of possible decoders. This set consists of all combinations of choices of $\hat{x}$ from the set $z$, for different pairs $(z,y)$. 

\subsubsection{The set of scalar source codes}
We define \emph{$\mathcal{F}^{WZ}(M)$} as the set of all scalar WZ source codes with rate $R=\log M$, i.e. all the pairs that consist of a scalar encoder and one of its possible decoders, as defined in (\ref{possible_decoders}).\\
$\mathbf{Remark}$. In contrast to our case, when there is a known joint distribution $P(x_i,y_i)$ , then given the encoder and $y_i$, the best strategy for minimizing the Hamming distortion is, of course, maximum likelihood, i.e., choose the most probable $x$ from the subset $m_{z_i}$, given $y_i$. 
\begin{equation}
         \hat{x}=\mbox{arg}\max_{x\in m_{z_i}}{P_{X|Y}(x|y_i)}    
\end{equation}
However, in our case, $P_{X|Y}(x|y)$ is unavailable since $P(x)$, the source statistics, is unknown or non-existent. Therefore, knowing the encoder is not sufficient for determining the best decoder.
\subsection{An on-line WZ coding scheme}
In this part, we describe an on-line adaptive scheme for the WZ case based on the results of [3]. For any source sequence $x^n$, the distortion $\Delta^n_{(e,d)}(x^n)$ of a source code $(e,d)$ is defined by:
\begin{equation}
\Delta^n_{(e,d)}(x^n)=\sum_{i=1}^n \Delta_i(x_i)
\label{xn_dis_def}
\end{equation}
where $\Delta_i(x_i)$ is as defined in (\ref{input_symbol_dis_def}).
In the case of a scalar source code, we get:
\begin{equation}
 \Delta^n_{(e,d)}(x^n)=\sum_{i=1}^n\sum_{y\in{\cal{X}}}P_{Y|X}(y|x_i)\rho(x_i,d(e(x_i),y))
\end{equation} 
Given any finite set of scalar source codes with rate $R$ and zero delay, this scheme (which has the same rate $R$ and zero delay) achieves asymptotically the distortion $\Delta^n_{(e,d)}(x^n)$ of the best source code in the set, for all source sequences $x^n$. 
To be more specific, we extend [3, Theorem 1] to include side information at the decoder. We get, that for any bounded distortion measure ($\rho(x,\hat{x})<B,\forall x,\hat{x}\in {\cal{X}}$ for some positive real number $B$), the following result holds:
\begin {theorem}
 Let $\mathcal{A}$ be a finite subset of $\mathcal{F}^{WZ}(M)$. Then there exists a sequential source code $(\tilde{e},\tilde{d})$ with rate $R=\log{M}$ and zero delay such that for all $x^n\in{\cal{X}}^n$:
\begin{equation}
\begin{array}{l}
\mathbb{E}\{\frac{1}{n}[\Delta^n_{(\tilde{e},\tilde{d})}({x^n})-\min_{(e',d')\in \mathcal{A}}\Delta^n_{(e',d')}(x^n)]\}\leq \frac{3B}{(2R)^{\frac{1}{3}}}[\log{|\mathcal{A}|}]^{\frac{2}{3}}n^{-\frac{1}{3}}
\end{array}
\end{equation}
\end {theorem}
where the expectation is taken w.r.t. a certain randomization of the algorithm, which will be described below.
For the Hamming distortion measure, we have $B=1$.
The proof is similar to the proof of [3, Theorem 1], where in our case, we use the distortion as defined in (\ref{xn_dis_def}). Since the proof steps are the same, we will not repeat them here.

The scheme works as follows: Assume some reference set $\mathcal{A}$ of WZ scalar source codes. We divide the time axis, $i= 1,2,\ldots,n$, into $K=n/l$ consecutive non-overlapping blocks (assuming $l$ divides $n$), where $l$ is a parameter to be determined.
At the beginning of each block, i.e., at times $t=(k-1)l,k\in\{1,2,\ldots,K\}$, we randomly choose an expert according to the exponential weighting probability distribution: 
\begin{equation}
Pr \{ next\ expert = (e,d) \} = \frac{\exp\{-\eta \Delta^t_{(e,d)}(x^t)\}}{\sum_{(e',d')\in \mathcal{A}}\exp\{-\eta \Delta^t_{(e',d')}(x^t)\}}
\label{expw_prob}
\end{equation}
where $\eta>0$ is a parameter to be determined. Notice that for $t=0$, we get uniform distribution.
After choosing the expert $(e,d)$, the encoder dedicates the first $\left\lceil \log|\mathcal{A}|/R\right\rceil$ channel symbols, at the beginning of the $k$-th block, to inform the decoder the identity of $d$. At the remainder of the block, the encoder produces the channel symbols $z_i=e(x_i)$. At the same time, at the decoder side, in the beginning of the block, at times $i=(k-1)l+1,\ldots,\left\lceil \log|\mathcal{A}|/R\right\rceil$, the decoder outputs arbitrary symbols from $\cal{X}$. At the rest of the block, knowing $d$, it reproduces $\hat{x}_i=d(z_i,y_i)$.

Exactly as in [3], the values of $l$ and $\eta$ are optimized to get minimal redundancy, and are given by:
\begin{equation}
\begin{array}{lll}
l&=&2\{\log(|\mathcal{A}|)n/R^2\}^{\frac{1}{3}}\\
\eta &=&\{8\log(|\mathcal{A}|)/lB^2n\}^{\frac {1}{2}}
\end{array}
\label{l_eta_opt}
\end{equation}
\\$\mathbf{Remark}$. Throughout this paper, we assume that $n$ is known in advance. Generalizing the scheme to the case where the horizon is unknown is straightforward, as explained in [3]. 
\section{Efficient implementation for sets of scalar source codes}
In this section, we present an efficient implementation of the scheme described in Section 2, for sets of scalar source codes. Each one of these sets of source codes consists of all pairs $(e,d)$, where $e$ is one of the encoders in some small set of encoders, and $d$ is one of its possible decoders, as defined in Subsection 2.1 . By ``small set", we mean that the random choice of the encoder can by done directly (as will be explained below). This definition depends, of course, on the computational resources we allocate.  
Remember that given a specific encoder, the decoder, for each $(z,y)$, chooses some $x$ from the subset of source letters $m_z$. Thus, for each pair $(z,y)$ there are $|m_z|$ possible $\hat{x}$'s. 
Hence, given an encoder, the number of possible decoders is:
\begin{equation}
\prod_{y\in{\cal{X}}}\prod_{z=1}^M{|m_z|}=(|m_1||m_2|\ldots|m_M|)^{|{\cal{X}}|}\geq 2^{|{\cal{X}}|}
\end{equation}
where $|m_z|$ is the cardinality of the subset of letters $m_z$.
The lower bound is derived from the fact that in the lossy encoding case $M<|{\cal{X}}|$, so the product above is at least $2$. Thus, given a set of encoders, the number of possible WZ source codes is at least $|E|•2^{|{\cal{X}}|}$, where $|E|$ is the number of encoders.
Given a set of experts $\mathcal{A}$, we follow the scheme of the previous subsection. We divide the time axis, $i= 1,2,\ldots,n$, into $K=n/l$ consecutive non-overlapping blocks. We randomly choose the next expert at the beginning of each block according to the exponential weighting probability distribution. The distortion of an expert $(e,d)$ at time $t$ is given by:
\begin{equation}
\begin{array}{lll}
\Delta^t_{(e,d)}(x^t)&=&\sum_{i=1}^t{\Delta(x_i)}\\&=&
\sum_{i=1}^t\sum_y{P_{Y|X}(y|x_i)}\rho(x_i,\hat{x}(x_i,y))\\&=&
\sum_{i=1}^t\sum_y{P_{Y|X}(y|x_i)I_{(x_i,y)\in A}}\\&=&
\sum_{x,y\in{\cal{X}}}n_t(x){P_{Y|X}(y|x)I_{(x_,y)\in A}}
\end{array}
\label{dis_def}
\end{equation} 
where $A$ is the set of all pairs $(x,y)$ which contribute to the distortion, i.e., $d(e(x),y)\neq x$, $I_B$ is the indicator function for an event $B$, and $n_t(x)$ is the number of times $x$ appeared in $x^t$.  
For a more convenient form of (\ref{expw_prob}), we multiply the numerator and denominator with $\exp\{\eta\sum_{x,y\in{\cal{X}}}n_t(x)P_{Y|X}(y|x)\}$ 
and we get:
\begin{equation}
Pr \{ next\ expert = (e,d) \} =\frac{\lambda_{(e,d),t}}{\sum_{(e',d')\in \mathcal{A}}\lambda_{(e',d'),t}}
\label{conv_form}
\end{equation}
where:
\begin{equation}
\lambda_{(e,d),t}=\exp\{\eta\sum_{x,y\in{\cal{X}}}n_t(x){P_{Y|X}(x,y)I_{(x,y)\in \bar{A}}}\}
\label{lamb_e_d_def}
\end{equation}
where $\bar{A}$ is the complementary set of $A$, i.e. all pairs $(x,y)$ such that $d(e(x),y)= x$.
Given a set of experts, the random choice of an expert at the beginning of each block is done in two steps. First, we choose an encoder randomly according to:
\begin {equation}
		Pr \{ next\ encoder = e \} = \frac{F_{e,t}}{\sum_{e'\in E}F_{e',t}}
\label{enc_choice}	
\end{equation}
where $E$ is the set of encoders, and: 
\begin {equation}
F_{e,t} = \sum_{(e,d)\in \mathcal{A}_e}\lambda_{(e,d),t}
\label{fe_def}
\end {equation}
is the sum of the exponential weights of all experts in $\mathcal{A}_e$, where $\mathcal{A}_e$ is the subset of all experts which use the encoder $e$. 
$F_{e,t}$ can be calculated efficiently in the following way:
For each pair $(x,y)$ calculate $\lambda_{x,y,t}$ where:
\begin{equation}
\lambda_{x,y,t}=\exp\{\eta n_t(x)P_{Y|X}(y|x)\}
\label{lamb_def}
\end{equation}
and then for each $(z,y)$, calculate the sum $\sum_{x:e(x)=z}\lambda_{x,y,t}$ where $e(x)$ is the 
encoding of $x$.
\begin{lemma} 
The product of all these sums is $F_{e,t}$:
\begin{equation}
F_{e,t}=\prod_{z=1}^M\prod_{y\in{\cal{X}}}\left(\sum_{x:e(x)=z}\lambda_{x,y,t}\right)
\label{fe_calc}
\end{equation}
\end{lemma}
$\mathbf{Proof}$.
\begin{equation}
\begin{array}{lll}
&&\prod_{z=1}^M\prod_{y \in \cal{X}}{(\sum_{x:e(x)=z}\lambda_{x,y,t})}\\\\&=&
\prod_{y \in \cal{X}}\prod_{z=1}^M{(\sum_{x:e(x)=z}\lambda_{x,y,t})}\\\\&=&
\prod_{y \in \cal{X}}\{\sum_{\overline{x}\in m_1\times m_2\ldots \times m_M}\prod_{i=1}^{M}\lambda_{\overline{x}(i),y,t}\}\\\\&=&
\sum_{\overline{x}_1,\overline{x}_2\ldots ,\overline{x}_{|\cal{X}|}\in m_1\times m_2\ldots \times m_M}\prod_{j=1 }^{|\cal{X}|}\prod_{i=1}^M\lambda_{\overline{x}_j(i),y_j,t}\\\\&=&
\sum_{\overline{x}_1,\overline{x}_2\ldots ,\overline{x}_{|\cal{X}|}\in m_1\times m_2\ldots \times m_{M}}\exp(\eta \sum_{j=1}^{|\cal{X}|}
\sum_{i=1}^{M} n_t(\overline{x}_j(i))P_{Y|X}(\overline{x}_j(i),y_j)\\\\&=&
\sum_{\overline{x}_1,\overline{x}_2\ldots ,\overline{x}_{|\cal{X}|}\in m_1\times m_2\ldots \times m_{M}}\exp(\eta \sum_{i,j=1}^{|\cal{X}|}
n_t(x_i)P_{Y|X}(x_i,y_j)\cdot I_{x_i \in (x_j(1),x_j(2)\ldots,x_j(|\cal{X}|))})\\\\&=&
\sum_{(e,d)\in \mathcal{A}_e}\lambda_{(e,d),t}
\end{array}
\label{proof_fe}
\end {equation}
In the second line, we change the order of the products, first over all $z$'s for a given $y$ and then on all the $y$'s.
In the third line, we calculate the product over $z$. $\overline{x}=(x(1),x(2),\ldots,x(M))$ is a vector of length $M$, where $\overline{x}(1)\in m_1$ (i.e. $e(\overline{x}(1))=1$), $\overline{x}(2)\in m_2$ etc.. The sum is over all such vectors. In other words, when expanding the product over $z$, we obtain the sum of all combinations of multiplying $M$ $\lambda_{x,y,t}$'s where the $x$ of each $\lambda$ belongs to a different subset of letters (according to the encoder). In the fourth line, we expand the product over $y$. Now, we obtain the sum of all combinations of multiplying $|\cal{X}|$ terms, where each product depends on some vector $\overline{x}$ as defined before, and on a different $y$. The rest is obtained by simply substituting the expression for $\lambda_{x,y,t}$. Choosing a decoder for a given encoder $e$, is actually choosing $|\cal{X}|$ vectors of length $M$, one vector for each $y$. The vector of each $y$, contains the decoding for each $z$ given that $y$, as explained for the third line.\\
In the second step, we choose the decoder randomly according to:
\begin {equation}
Pr\{decoder =d\ |\ encoder=e\} = \frac{\lambda_{(e,d),t}}{F_{e,t}}
\label{dec_choice}
\end {equation}
The random choice of the decoder can be implemented efficiently in the following way:
For each pair $(z,y)$, choose the decoder output $d(z,y)$ randomly, according to the probability distribution:
\begin {equation}
Pr\{d(z,y)=x\}= \frac{\lambda_{x,y,t}}{\sum_{x':e(x')=z}\lambda_{x',y,t}}
\label{dec_calc}
\end {equation}
where $x\in \{ x : e(x)= z \}$.
Choosing the decoder function in this way, we get that:
\begin{equation}
Pr\{decoder =d\ |\ encoder=e\}=\\
\prod_{y\in \cal{X}}\prod_{z=1}^M Pr\{d(z,y)=x\}=\\
\frac{\prod_{y\in \cal{X}}\prod_{z=1}^M\lambda_{x,y,t}}{\prod_{y\in \cal{X}}\prod_{z=1}^M\sum_{x':e(x')=z}\lambda_{x',y,t}}
\end{equation}
The numerator and denominator were already proved to be given by $\lambda_{(e,d),t}$ and $F_{e,t}$, respectively in (\ref{proof_fe}). Therefore, the decoder is indeed chosen according to (\ref{dec_choice}).
We demonstrated an efficient random selection of a pair $(e,d)$.
Below is a formal description of the on-line algorithm:
\begin{enumerate}
	\item Calculate $l$, the optimal length of a data block, according to (\ref{l_eta_opt}), and let $K = n / l$.
	\item Initialize $k$ to $0$, and all the weights $\lambda_{x,y,0}$ to 1.
	\item At the beginning of block no. $k$, update the weights in the following way:\\
	$\lambda_{x,y,t_k} = \lambda_{x,y,t_{k-1}}\exp(\eta  \sum_{i=(k-1)•l+1}^{k•l}I_{x_i=x}  P_{Y|X}(x,y) )
	\\t_k = kl+1,\ 1\leq k \leq K-1$
	\item For each $(e,z,y)$, calculate the sums:\\
	 $\sum_{x: e(x)=z}\lambda_{x,y,t_k}$
	 \item Calculate $F_{e,t_k}$, for each $e\in E$, according to (\ref{fe_calc}).
	 \item Choose an encoder $e_k$ randomly according to (\ref{enc_choice}).
   \item For each pair $(z,y)$, choose the decoder function $d_k$ randomly according to (\ref{dec_calc}).
   \item Use the first $\left\lceil \log(N)/R\right\rceil$ channel symbols at the beginning of the $k$th block to inform the decoder the identity of $d_k$, chosen in the previous step, where $N$ is the number of experts.
   \item Encode the next block using the chosen expert $e_k$:\\
   $z_i= e_k(x_i),\ k•l+\log(N)/R+1 \leq i \leq (k+1)•l-1$
   \item If $k<K$, increment $k$ and go to 3.  
\end{enumerate}
The total complexity of the algorithm is $O(n/l\cdot|{\cal{X}}|^2)+O(n/l\cdot|E||{\cal{X}}|M)+O(n)$. The complexity depends on $|E|$, which thus should be small as was mentioned above.\\
The computational complexity of the algorithm is as follows: The calculations of $\sum_{i=(k-1)•l+1}^{k•l}I_{x_i=x}$ for each $x\in{\cal{X}}$ at each time $t_k$ take $O(n)$ computations totally. After calculating these quantities, we can update the $\lambda_{x,y,t}$'s as described in step 3 of the algorithm above. This takes $O(|{\cal{X}}|^2)$. Calculating the sums in step 4, given the $\lambda_{x,y,t}$'s, takes $O(|{\cal{X}}|^2)$. Calculating the $F_{e,t}$'s takes $O(|E||{\cal{X}}|M)$ calculations where $|E|$ is the cardinality of the set of encoders.
\subsection{Large set of encoders with structure}
As was shown, we can choose a pair $(e,d)$ randomly, in two steps. In the first step, we choose the encoder according to (\ref{enc_choice}). In the second step, we choose randomly one of its possible decoders according to (\ref{dec_choice}). In the previous part, we assumed that the set of encoders is small, so we can implement (\ref{enc_choice}) directly, i.e., calculate $F_{e,t}$ for each encoder separately. In this part, we use a large structured set of encoders. Using the structure, we can efficiently implement (\ref{enc_choice}). We assume that the input alphabet ${\cal{X}}$ is ordered. We enumerate the source symbols according to that order. By $Num(x),1\leq Num(x)\leq|{\cal{X}}|$, we denote the location of the symbol $x$ in that order.
\subsubsection{Definition of the set of encoders}
The Input Alphabet Axis (IAA) is defined as the $|{\cal{X}}|$-dimensional vector $(1, 2, \ldots, |{\cal{X}}|)$. A partition of the IAA is given by the $(M-1)$-dimensional sequence $\mathbf{r}=(z_1,\ldots,z_{M-1}), z_i\in \{1,2,\ldots,|{\cal{X}}|-1\}, 0\equiv z_0<z_1<\ldots<z_M\equiv |\cal{X}|$. Each partition $r$ represents a specific encoder in the following way:
\begin{equation}
\begin {array} {ll}
e(x)= i:z_{i-1} < Num(x) \leq z_i,&i\in \{1,\ldots,M\}
\end{array}
\end{equation}We define $E$ as the set of all such encoders.
The cardinality of the set of encoders is $\left(\begin{array}{c}|{\cal{X}}|\\M-1\end{array}\right)$. 
\subsubsection{Graphical representation of the set of encoders}
The random choice of the encoders can be done efficiently using an a-cyclic directed graph (see Fig. 1).
We use the following notation:\\
$\mathcal{V}$ - The set of all vertices:\\
$\{1, 2,\ldots ,|{\cal{X}}|-1\}\times \{1, 2, \ldots ,M-1\} \cup (0,0) \cup (|{\cal{X}}|,M)$\\
$\mathcal{E}$ - The set of all edges:\\
$\{ ( (z,j-1),(\hat{z},j)) : z,\hat{z}\in\{0,1,2,\ldots ,|{\cal{X}}|\}, j\in\{1,2,\ldots ,M\},\hat{z}>z\}$\\
$s$ - The starting point in the bottom left, i.e. $(0,0)$\\
$u$ - The end point in the top right, i.e. $(M,|{\cal{X}}|)$\\
$\mathcal{E}_z$ - The set of all edges starting from vertex $z$.\\
A general graph is described in Fig. 1.
The horizontal axis represents the ordered input alphabet.
The vertical axis represents the $M-1$ choices needed for dividing the IAA into $M$ segments.
A path composed of the edges $\{ (0,0), (z_1,1)\ldots , (z_{M-1},M-1),(|{\cal{X}}|,M)\}$ represents $M-1$ consecutive choices of $M-1$ $x$'s $(z_1,\ldots,z_{M-1})$ which divide the IAA into $M$ segments, creating $M$ subsets of the input alphabet.
Each edge on a path represents one choice, the choice of the next point on the horizontal axis, which defines the next segment. 
An edge $((z,j-1),(\hat{z},j))$ matches to the segment $(z,\hat{z}]$ on the horizontal axis, thus equivalent to the subset $\{x:z < x \leq \hat{z}\}$. 
There are $O(M|{\cal{X}}|^2)$ edges.
\begin{figure}[hc]
	\centering		\includegraphics[width=5in, height=3in]{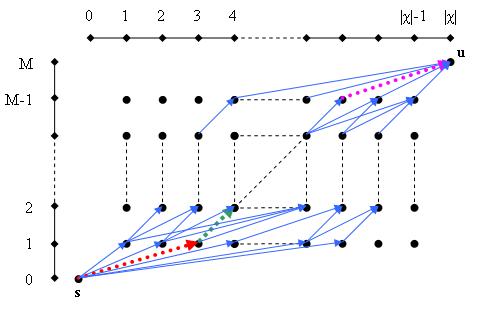} 
	\caption { {The graph representing all possible partitions of the input alphabet into $M$ subsets given the alphabet is ordered in some specific order. For example, the left dashed arrow defines the subset $\{1,2,3\}$, the middle and right dashed arrows define the subsets $\{4\}$ and $\{|{\cal{X}}|-2,|{\cal{X}}|-1,|{\cal{X}}|\}$ respectively.}}
\end{figure}
For each edge $a\in\mathcal{E}$ and time $t$ we assign a weight $\delta_{a,t}$:
\begin{equation}
\begin{array}{ll}
\delta_{a,t}=\prod_{y\in{\cal{X}}}\sum_{x\in(z,\hat{z}]}\lambda_{x,y,t} ,&a=((z,j-1),(\hat{z},j))
\end{array}
\label{lam_wei}
\end{equation}
where $\lambda_{x,y,t}$ is given by (\ref{lamb_def}). It can be seen from (\ref{lam_wei}) that a weight $\delta_{a,t}$ depends only on the horizontal coordinates of the edge $a$, thus we can denote it as $\delta_{(z,\hat{z}),t}$.
The cumulative weight of a path $r=\{ (0,0), (z_1,1)\ldots , (z_{M-1},M-1),(|{\cal{X}}|,M)\}$ at time $t$ is defined as the product of its edge weights:
\begin{equation}
\Lambda_{r,t}=\prod_{a\in r} \delta_{a,t}
\end{equation}
$\Lambda_{r,t}$ is simply $F_{e,t}$:
\begin{equation}
\Lambda_{r,t}=\prod_{a\in r} \delta_{a,t}=\prod_{m=1}^M\prod_{y\in\cal{X}}\sum_{x\in(z_{m-1},z_m]}\lambda_{x,y,t}=F_{e,t}
\label{fe_is_product}
\end{equation}
where the last equality was proved in (\ref{proof_fe}).
From now on, our WPA description is general, thus will be used for all a-cyclic directed graphs in this article.
Following the WPA, also used in [4] and [5] we define:
\begin {equation}
G_t(z)=\sum_{r\in \mathcal{R}_z} \Lambda_{r,t}
\end{equation}
where now, $z$ is a vertex on the graph (and not only coordinate), $\mathcal{R}_z$ is the set of all paths from $z$ to $u$ and $a$ is an edge on the path $r$.\\
We see that:
\begin{equation}
G_t(s)=\sum_{r\in \mathcal{R}_s} \Lambda_{r,t}=\sum_{e\in E}F_{e,t}
\end{equation}
where $E$, the set of encoders, is of course equivalent to ${\cal{R}}_s$, the set of all paths from $s$ to $u$.
The function $G_t(z)$ can be computed recursively:
\begin{equation}
\begin {array}{l}
G_t(u)=1,\ \ G_t(z) = \sum_{\hat{z}:(z,\hat{z})\in \mathcal{E}}\delta_{(z,\hat{z}),t}G_t(\hat{z})
\end{array}
\label{g_t_recursive}
\end{equation}
Because each edge is taken exactly once, calculating $G_t(z)$ for all $z$'s requires $O(|\mathcal{E}|)$ computations given the weights $\delta_{a,t}$.
The function $G_t(z)$ offers an efficient way to choose an encoder randomly according to probability distribution in (\ref{enc_choice}).\\
We define for each $\hat{z}\in \mathcal{E}_z$: 
\begin{equation}
P_t(\hat{z}|z) =\delta_{(z,\hat{z}),t}G_{t}(\hat{z})/G_{t}(z)
\label{seq_decis_prob}
\end{equation}
It is easy to see that $P_t(\hat{z}|z)$ is indeed a probability distribution, i.e., $\sum_{\hat{z}:(z,\hat{z})\in \mathcal{E}_z} P_t(\hat{z}|z) = 1$. 
We also have: 
\begin{eqnarray}
\nonumber\prod_{m=1}^{M}P_t(z_{m,r}|z_{m-1,r})&=&\nonumber\prod_{m=1}^{M}\delta_{(z_{m-1,r},z_{m,r}),t}\frac{G_{t}(z_{m,r})}{G_{t}(z_{m-1,r})}\\
&=&\nonumber\frac{G_{t}(u)}{G_{t}(s)}\cdot\prod_{m=1}^{M}\delta_(z_{m-1,r},z_{m,r})\\
&=&Pr\{ next\ encoder = r\}
\end{eqnarray}
and we get exactly the probability in (\ref{enc_choice}).\\ 
Therefore, the encoder can be chosen randomly in the following sequential manner: Starting from $z_0=s$, at each step $m = 1, 2, \ldots,M-1$, choose the next vertex $z_m\in \mathcal{E}_{z_{m-1}}$ with probability $P_t(z_m|z_{m-1})$. The procedure stops when $z_m=u$.\\

\noindent\emph{Formal description of the on-line algorithm}: Using the set of encoders described above, we now have the following algorithm:
\begin{enumerate}
	\item Calculate $l$, the optimal length of a data block, according to (\ref{l_eta_opt}), and let $K = n / l$.
	\item Initialize $k$ to $0$, and all the weights $\lambda_{x,y,0}$ to 1.
	\item Build the encoders graph as described in this section.
	\item Initialize all the weights $\delta_{a,0}$ to 1.
	\item At the beginning of block no. $k$, i.e. at time $t_k=kl+1, k=\{1,\ldots,K\}$  update the weights in the following way:\\
	$\lambda_{x,y,t_k} = \lambda_{x,y,t_{k-1}}\exp(\eta  \sum_{i=(k-1)•l+1}^{k•l}I_{x_i=x}  P_{Y|X}(x,y) )$
	\item At the beginning of block no. $k$, calculate $\delta_{z,\hat{z},t_k}$ for each pair $(z,\hat{z})$ according to (\ref{lam_wei}).
	\item Update the weights of all edges to the new $\delta_{(z,\hat{z}),t_k}$'s.
	 \item Calculate $G_{t_k}(z)$ recursively, for all $z$, according to (\ref{g_t_recursive}).
	 \item Choose the encoder $e_k$ randomly as described above, using (\ref{seq_decis_prob}).
    \item For each pair $(z,y)$, choose the decoder function $d_k$ randomly according to (\ref{dec_calc}).
   \item Use the first $\left\lceil \log(N)/R\right\rceil$ channel symbols at the beginning of the $k$th block to inform the decoder the identity of $d_k$, chosen in the previous step, where $N$ is the number of experts.
   \item Encode the next block, using the chosen expert $e_k$:\\
   $z_i= e_k(x_i),\ k•l+\log(N)/R+1 \leq i \leq (k+1)•l-1$
   \item If $k<K$, increment $k$ and go to 3.     
\end{enumerate}
The total complexity of the algorithm is $O(n/l\cdot|{\cal{X}}|^3)+O(n/l\cdot M|{\cal{X}}|^2)+O(n/l|{\cal{X}}|^2)+O(n)$.
\section{Definition of an On-line Adaptive Variable-Rate Coding Scheme}
In this section, we generalize the results of Theorem 1 to the variable-rate coding case. This is done by generalizing the performance criteria, to include also the compression, in addition to the cumulative distortion of the code. The scheme we use is similar to that of theorem 1.
The use of variable-rate codes complicates the problem. A choice of an expert is actually a combination of two choices. We now have to choose simultaneously, a lossy code and a lossless variable-rate code (as will be explained).
The challenge is to describe the reference set in such a way that allow us efficient implementation.
We start by defining a variable-rate code. Without loss of generality, we assume that the compressed sequence is binary.
We define $C_M$ as a binary prefix code, which contains $M$ codewords $\{b_1, b_2, \ldots, b_M\}$, where each $b_i$ is a binary string of length $l(b_i)$. We call the $\mathbf{ordered}$ set $\{l(b_1), l(b_2),\ldots, l(b_M)\}$, the $length\ set$ of the code. Since we deal with prefix codes, a length set must, of course, maintain the Kraft inequality, i.e., $\sum_{i=1}^M 2^{-l(b_i)}\leq 1$.
A source code of variable-rate is defined in the following way:
Given a source sequence $x^n$, the operation of the encoder can be described as being composed of two steps, the first one is lossy and the second is lossless.
First, the encoder transforms $x^n$ into a sequence $z^n$ whose symbols $\{z_i\}$ take on values in the set $\{1,2,\ldots,M\}$. 
If $M<\log|{\cal{X}}|$, this step is of course lossy. Then, the encoder uses a code $C_M$ to encode $z^n$ into $b^n$, a sequence of variable-length binary strings, by encoding each $z_i$ into a codeword $b_i$, where $b_i$ takes on values in a set $\{b_1,b_2,\ldots,b_M\}$. This step is lossless. The decoder, knowing $b^n$, produces $z^n$ without error. Then, based on $z^n$ and the side information sequence $y^n$, the decoder produces the reconstructed sequence $\hat{x}^n$. Throughout the rest of the paper, we omit the intermediate sequence $z^n$, and the lossless part of the decoding. 
We define the Lagrangian Cost (LC) for the input symbol at time $t$, $x_t$, as:
\begin{equation}
{\cal{L}}(x_t)=\Delta(x_t)+ \delta l(b_t)
\end{equation}
where $\Delta(x_t)\in[0,B]$ is a bounded distortion measure, $l(b_t)$ is the length of the binary codeword at time $t$ and $\delta$ is a positive constant. If the $C_M$'s are Huffman codes, $l(b_t)$ is bounded by $M-1$, the maximal depth of a complete binary tree with $M$ leaves.

\subsection{Definition of the reference set of source codes}
In this part, we define the general set of variable-rate scalar source codes, i.e., our set of experts.
Each expert is a source code with $M$ binary codewords which consists of some binary prefix code $C_M$. Each expert partitions $\cal{X}$ into $M$ disjoint subsets $( m_1, m_2,\ldots, m_M)$, where each subset $m_i$ is encoded as a binary codeword $b_i$, $i\in\{1,2,\ldots,M\}$. The variable-rate encoder $e$ for each expert is given by a function $e : \cal{X} \rightarrow$$\{b_1, b_2, \ldots, b_M\}$ that is, $b_i=e(x_i)$.
The decoder $d$ receives $b_i$, and together with the side information $y_i$ if available, decides on $\hat{x}_i$, using a decoding function 
$d : \{b_1, b_2, \ldots,b_M\}\times\cal{X}\rightarrow\cal{X}$, i.e., $\hat{x}_i=d(b_i,y_i)$. 
The set of decoders is defined as in Subsection 2.1, with only one difference: instead of getting an index $z_i$, it gets a binary codeword which represents this index. Again, we limit our discussion to decoders which satisfy:
\begin{equation} 
d(e(\hat{x}),y)= \hat{x}, for\ all\ \hat{x}\ and\ y 
\end{equation}
To complete the definition, as was explained in Subsection 2.1, all the encoders which have the same functionality is treated as the same encoder. In this part, by $same\ functionality$, we mean that encoders which implement the same partition of the input alphabet as was defined in Subsection 2.1, and in addition, have the same length set, are treated as the same encoder.

We define \emph{$\mathcal{G}^{VR}(M)$} as the set of all variable-rate scalar source codes, i.e., all the pairs of variable-rate scalar encoders and one of their possible decoders, as defined in this section.

\subsection{An on-line variable-rate coding scheme}
In this part, we describe an on-line adaptive variable-rate scheme coding based on the results of [3]. We actually extend Theorem 1
from the case of a pure distortion criterion to the LC case.
For any source sequence $x^n$, the LC ${\cal{L}}^n_{(e,d)}(x^n)$ of a source code $(e,d)$ is defined by:
\begin{equation}
{\cal{L}}^n_{(e,d)}(x^n)=\sum_{t=1}^n {\cal{L}}(x_t)
\end{equation}
For the WZ case of scalar source code we get:
\begin{equation} 
{\cal{L}}^n_{(e,d)}(x^n)=\sum_{t=1}^n\sum_{y\in{\cal{X}}}P_{Y|X}(y|x_t)\rho(x_t,d(e(x_t),y))+\delta \sum_{t=1}^n l(b_t).
\end{equation}
Given any finite set of variable-rate scalar source codes, our coding scheme achieves asymptotically the LC, ${\cal{L}}^n_{(e,d)}(x^n)$, of the best source code in the set, for all source sequences $x^n$.
To be more specific, it can be shown, in a similar way as in [3], that for any bounded distortion measure and some positive $\delta$, the following result holds:
\begin {theorem}
 Let $\mathcal{A}$ be a finite subset of $\mathcal{G}^{VR}(M)$. Then there exists a sequential source code $(\tilde{e},\tilde{d})$ such that for all $x^n\in{\cal{X}}^n$:
\begin{equation}
\begin{array}{l}
\mathbb{E}\{\frac{1}{n}[{\cal{L}}^n_{(\tilde{e},\tilde{d})}({x^n})-\min_{(e',d')\in \mathcal{A}}{\cal{L}}^n_{(e',d')}(x^n)]\}\leq C_1[\log{|\mathcal{A}|}]^{\frac{2}{3}}n^{-\frac{1}{3}}
\end{array}
\end{equation}
Where $C_1$ is a constant depends only on $B$, $M$ and $\delta$.
\end {theorem}
The proof is similar to the proof of Theorem 1. Nonetheless, we give the full proof for completeness because there are some differences between the case of fixed-rate and the variable-rate case.\\
$\mathbf{Proof\ of\ Theorem\ 2:}$\\
The scheme works similarly to the scheme in the fixed-rate case: Assume that we have some reference set $\mathcal{A}$ of variable-rate WZ scalar source codes. We divide the time axis, $t= 1,2,\ldots,n$, into $K=n/l$ consecutive non-overlapping blocks (assuming $l$ divides $n$).
At the beginning of each block, i.e., at time $t=(k-1)l,k\in\{1,2,\ldots,K\}$, we randomly choose an expert according to the exponential weighting probability distribution: 
\begin{equation}
Pr \{ next\ expert = (e,d) \} = \frac{\exp\{-\eta {\cal{L}}^t_{(e,d)}(x^t)\}}{\sum_{(e',d')\in \mathcal{A}}\exp\{-\eta {\cal{L}}^t_{(e',d')}(x^t)\}}
\label{expw_prob2}
\end{equation}
The parameters $l$ and $\eta>0$ will be optimized later . 
After choosing the expert $(e,d)$, the encoder dedicates the first $\left\lceil \log|\mathcal{A}|\right\rceil$ bits, at the beginning of the $k$-th block, to inform the decoder the identity of $d$.
At the remainder of the block, the encoder produces the binary strings $b_i=e(x_i)$. At the same time, at the decoder side, when getting the first  $\left\lceil \log|\mathcal{A}|\right\rceil$ bits of the block, the decoder outputs arbitrary symbols. At the rest of the block, knowing $d$, it reproduces $\hat{x}_i=d(b_i,y_i)$.
Define for each $k$:
\begin{equation}
W_k=\sum_{(e',d')\in \mathcal{A}}\exp\{-\eta {\cal{L}}_{(e',d')}^{(k-1)l}(\mathbf{x})\}
\end{equation}
As in [3], we then have for $N=n/l$:
\begin{eqnarray}
\nonumber\log\frac{W_{N+1}}{W_1}&=&\log\sum_{(e',d')\in \mathcal{A}}\exp\{-\eta {\cal{L}}_{(e',d')}^{Nl}(\mathbf{x})\}-\log|\mathcal{A}|\\
\nonumber&\geq&\log \max_{(e',d')\in \mathcal{A}}\exp\{-\eta {\cal{L}}_{(e',d')}^{Nl}(\mathbf{x})\}-\log|\mathcal{A}|\\
&=&-\eta \min_{(e',d')\in \mathcal{A}}{\cal{L}}_{(e',d')}^{n}(\mathbf{x})-\log|\mathcal{A}|
\label{proof21}
\end{eqnarray}
On the other hand, for each $1\leq k\leq n$
\begin{eqnarray}
\nonumber\log\frac{W_{k+1}}{W_k}&=&\log\frac{\sum_{(e',d')\in \mathcal{A}}\exp\{-\eta {\cal{L}}^{(k-1)l+1,kl}_{(e',d')}(\mathbf{x})\}\exp\{-\eta {\cal{L}}^{(k-1)l}_{(e',d')}(\mathbf{x})\}}{\sum_{(e',d')\in \mathcal{A}}\exp\{-\eta {\cal{L}}^{(k-1)l}_{(e',d')}(\mathbf{x})\}}\\
\nonumber &=&\log{\mathbb{E}_{Q_k}}\exp\{-\eta {\cal{L}}^{(k-1)l+1,kl}_{(e',d')}(\mathbf{x})\}\\
\nonumber &\leq& -\eta{\mathbb{E}_{Q_k}} {\cal{L}}^{(k-1)l+1,kl}_{(e',d')}(\mathbf{x})+\frac{{\eta}^2 l^2 \tilde{B}^2}{8}\\
&\leq& -\eta \{\mathbb{E}[{\cal{L}}^{(k-1)l+1,kl}_{(e,d)}(\mathbf{x})]-B\log|\mathcal{A}|\}+\frac{{\eta}^2 l^2 \tilde{B}^2}{8}
\end{eqnarray}
where in our case, we have defined the maximum LC for a single input symbol:
\begin{equation}
\tilde{B}=B+\delta(M-1)
\end{equation}
and where ${\mathbb{E}_{Q_k}}$ denotes expectation with respect to the distribution $Q_k$ on $\mathcal{A}$, which assigns a probability proportional to $\exp\{-\eta {\cal{L}}^{(k-1)l}_{(e',d')}\}$ to each $(e',d')$ in $\mathcal{A}$. The expectation in the last line is with respect to the random choices of the code.
The first inequality follows from the Hoeffding's bound (cf. [11, Lemma 8.1]). The second follows from the construction of the code described above. We use the first $\left\lceil \log|\mathcal{A}| \right\rceil$ bits of each data block, to inform the decoder the identity of the encoder. This causes a cumulative distortion which depends on the number of codewords we lose. Unlike in [3], this number is not constant because we use a variable-length code. The maximal number of codewords we can lose is $\left\lceil \log|\mathcal{A}| \right\rceil -1$ (The worst case is when each one of the first $\left\lceil \log|\mathcal{A}| \right\rceil -1$ codewords have length of one bit. Since the $\left\lceil \log|\mathcal{A}| \right\rceil$ bit belongs to the next codeword we lose it too). Therefore, the cumulative distortion caused by the lose of the first $\left\lceil \log|\mathcal{A}| \right\rceil$ bits can be no more than $B\left\lceil \log|\mathcal{A}| \right\rceil$. The cumulative distortion of the rest of the block is exactly the loss of the pair $(e,d)$ chosen at the beginning of the block. 
Summing over $k$, we get:
\begin {equation}
\log\frac{W_{N+1}}{W_1} \leq -\eta \sum_{k=1}^N\mathbb{E}[{\cal{L}}^{(k-1)l+1,kl}_{(e',d')}(\mathbf{x})]+\eta BN\log|\mathcal{A}|+\frac{{\eta}^2 l^2 \tilde{B}^2 N}{8}
\label{proof22}
\end {equation}
Combining (\ref{proof21}) and (\ref{proof22}), we get:
\begin{eqnarray}
\nonumber && \sum_{k=1}^N\mathbb{E}[{\cal{L}}^{(k-1)l+1,kl}_{(e',d')}(\mathbf{x})]-\min_{(e',d')\in \mathcal{A}}{\cal{L}}^n_{(e',d')}(\mathbf{x})\\
\nonumber &\leq& \frac{\log|\mathcal{A}|}{\eta}+\frac{\eta l^2 \tilde{B}^2 N}{8}+BN\log|\mathcal{A}|\\
&=&\tilde{B}\sqrt{\log|\mathcal{A}|/2}\cdot l \cdot N^{\frac{1}{2}}+BN\log|\mathcal{A}|
\label{optim}
\end{eqnarray}
where the equality follows upon taking the minimizing value $\eta=\sqrt{8\log|\mathcal{A}|/l^2 \tilde{B}^2 N}$.
For convenience, we denote $\alpha=\tilde{B}\sqrt{\log|\mathcal{A}|/2}$ and $\beta=B\log|\mathcal{A}|$, so that the last line of (\ref{optim}) becomes $\alpha l N^{\frac{1}{2}}+\beta N=\alpha nN^{-\frac{1}{2}}+\beta N$.
Minimizing with respect to $N$ we take $N=(\alpha n /2\beta)^{\frac{2}{3}}$ and get an expression upper bounded by $2(\alpha n)^{\frac{2}{3}}\beta^{\frac{1}{3}}$. Placing the values of $\alpha,\beta$ we obtain:
\begin {equation}
\mathbb{E}\{\frac{1}{n}[{\cal{L}}^n_{(\tilde{e},\tilde{d})}(\mathbf{x})-\min_{(e',d')\in \mathcal{A}}{\cal{L}}^n_{(e',d')}(\mathbf{x})]\}\leq B^{\frac{1}{3}}(2\tilde{B}\log|\mathcal{A}|)^{\frac{2}{3}}n^{-\frac{1}{3}}
\end {equation}
Throughout the above proof, we got the following optimized values for $l$ and $\eta$:
\begin{equation}
\begin{array}{lll}
l&=&2\{\log(|\mathcal{A}|)nB^2/ \tilde{B}^2\}^{\frac{1}{3}}\\
\eta&=&\{8\log|\mathcal{A}|/l \tilde{B}^2 n\}^\frac{1}{2}
\end{array}
\label{l_eta_def_2}
\end{equation}
\section{Efficient implementation for sets of scalar source codes with variable-rate coding}
In this section, we present an efficient implementation of the scheme described, for sets of variable-rate scalar source codes. Each one of these sets of source codes consists of all pairs $(e,d)$ where $e$ is one of the encoders in some small set of encoders, and $d$ is one of its possible decoders. At the beginning of Section 3, $E$ was defined as some small set of fixed-rate encoders. We now define ${\mathcal{H}}$ as a small set of binary prefix codes. In this section, our set of encoders is defined to be $E\times{\cal{H}}$, which is all the encoders obtained by a combination between one of the encoders of $E$ and a binary prefix code belongs to $\mathcal{H}$. Generalizing (\ref{dis_def}), the LC of an expert $(e,d)$ at time $t$ is given by:
\begin{equation}
{\cal{L}}^t_{(e,d)}(x^t)=\sum_{x,y\in{\cal{X}}}n_t(x){P_{Y|X}(y|x)I_{(x_,y)\in A}}+\delta\sum_{x\in{\cal{X}}}n_t(x)l(e(x))
\end{equation}
where $A$, $n_t(x)$ and $I_B$ were defined in (\ref{dis_def}) and $l(e(x))$ is the length of the codeword $b=e(x)$. As in (\ref{conv_form}), we change to a more convenient form by multiplying the numerator and denominator by $\exp\{\eta\sum_{x,y\in{\cal{X}}}n_t(x)P_{Y|X}(y|x)\}$, and we get:
\begin{equation}
Pr \{ next\ expert = (e,d) \} =\frac{\lambda_{(e,d),t}\cdot \gamma_{e,t}}{\sum_{(e',d')\in \mathcal{A}}\lambda_{(e',d'),t}\cdot \gamma_{e',t}}
\end{equation}
where $\lambda_{(e,d),t}$ is given by (\ref{lamb_e_d_def}) and where we define:
\begin{equation}
\gamma_{e,t}=\exp\{-\eta\cdot\delta\sum_{x\in{\cal{X}}}n_t(x)l(e(x))\}
\end{equation}
As in Section 3, given a set of experts, the random choice of an expert at the beginning of each block is done in two steps. First, we choose an encoder randomly according to:
\begin {equation}
		Pr \{ next\ encoder = e \} = \frac{F^{\cal{L}}_{e,t}}{\sum_{e'\in E\times {\cal{H}}}F^{\cal{L}}_{e',t'}}
\label{enc_choice2}
\end{equation}
where $E\times {\mathcal{H}}$ is the set of encoders, and: 
\begin {equation}
\begin{array}{lll}
F^{\cal{L}}_{e,t} &=&\sum_{(e,d)\in \mathcal{A}_e}\exp\{-\eta{\cal{L}}^t_{(e,d)}(x^t)\}\\&=& \sum_{(e,d)\in \mathcal{A}_e}\lambda_{(e,d),t}\cdot\gamma_{e,t}\\&=&
\gamma_{e,t}\cdot\sum_{(e,d)\in \mathcal{A}_e}\lambda_{(e,d),t}\\&=&
\gamma_{e,t}\cdot F_{e,t}
\end{array}
\label{f_e_lc_calc}
\end {equation}
is the sum of the exponential weights of all experts in $\mathcal{A}_e$, where $\mathcal{A}_e$ is the subset of all experts which use the encoder $e$, and $F_e$ was defined in (\ref{fe_def}). It was shown in Section 3 that $F_e$ can be calculated efficiently. $\gamma_{e,t}$ can be calculated directly for each $e$, given that $|E\times{\cal{H}}|$ is reasonably small.
In the second step, we choose the decoder randomly exactly as we did before, according to (\ref{dec_choice}).
Let us show that the pair $(e,d)$ is indeed chosen according to (\ref{enc_choice2}):
\begin{equation}
\begin{array}{lll}
Pr \{ next\ expert = (e,d) \}&=&Pr \{ next\ encoder = e \}\cdot Pr\{decoder =d\ |\ encoder=e\}\\&=&
({F^{\cal{L}}_{e,t}}/{\sum_{e'\in E\times {\cal{H}}}F^{\cal{L}}_{e',t'}})\cdot({\lambda_{(e,d),t}}/{F_{e,t}})\\&=&
\gamma_{e,t}\cdot\lambda_{(e,d),t}/{\sum_{e'\in E\times {\cal{H}}}F^{\cal{L}}_{e',t'}}
\end{array}
\end{equation}
We demonstrated an efficient random choice of a pair $(e,d)$.
Below is a formal description of the on-line algorithm:
\begin{enumerate}
	\item Calculate $l$, the optimal length of a data block, according to (\ref{l_eta_def_2}), and let $K = n / l$.
	\item Initialize $k$ to $0$, and all the weights $\lambda_{x,y,0}$ and $\gamma_{e,0}$ to $1$.
	\item At the beginning of block no. $k$, update the weights in the following way:\\
	$\begin{array}{lll}
	\lambda_{x,y,t_k} &=& \lambda_{x,y,t_{k-1}}\exp\{\eta  \sum_{i=(k-1)•l+1}^{k•l}I_{x_i=x}  P_{Y|X}(x,y) \}
	\\t_k &=& kl+1,\ 1\leq k \leq K-1
	\end{array}$
	\item For each $(e,z,y)$, calculate the sums:\\
	 $\sum_{x: e(x)=z}\lambda_{x,y,t_k}$
	 \item Calculate $F_{e,t_k}$, for each $e\in E$, according to (\ref{fe_calc}).
	 \item Update the $\gamma$'s, for each $e\in E\times{\cal{H}}$, in the following way:\\
	 $\gamma_{e,t_k}=\gamma_{e,t_{k-1}}\exp\{-\eta\cdot\delta  \sum_{i=(k-1)•l+1}^{k•l} l(e(x_i)) \}$
	 \item Calculate $F^{\cal{L}}_{e,t_k}$, for each $e\in E\times{\cal{H}}$ according to (\ref{f_e_lc_calc}).
	 \item Choose an encoder $e_k$ randomly according to (\ref{enc_choice2}).
   \item For each pair $(z,y)$, choose the decoder function $d_k$ randomly according to (\ref{dec_calc}).
   \item Use the first $\left\lceil \log(N)\right\rceil$ bits at the beginning of the $k$th block to inform the decoder the identity of $d_k$, chosen in the previous step, where $N$ is the number of experts.
   \item Encode the next block using the chosen expert $e_k$:\\
   $b_i= e_k(x_i),\ k•l+\log(N)+1 \leq i \leq (k+1)•l-1$
   \item If $k<K$, increment $k$ and go to 3.  
\end{enumerate}
Notice that in step 10, the lower bound on $i$ is the worst case, as was explained in the proof of Theorem 2.
The total complexity of the algorithm is $O(n/l\cdot|{\cal{X}}|^2)+O(n/l\cdot|E\times{\cal{H}}|\cdot M)+O(n/l\cdot|E||{\cal{X}}|M)+O(n)$. The complexity depends on $|E\times{\cal{H}|}$, which thus should be small as was mentioned above.\\
In the following subsection, we first show an efficient scheme which use an a-cyclic directed graph and the WPA to implement an adaptive Huffman coding. We then use the idea of representing all Huffman codes by a graph, to extend the scheme for structured sets of encoders from Subsection 3.1, and build a full LC scheme for the WZ case.
\subsection{An efficient adaptive lossless coding scheme using Huffman codes}
In this subsection, we assume that the input alphabet have $M$ symbols and we use $M$ codewords, thus the encoding is lossless. This is a special case of the general LC coding we defined in the previous parts, when $\Delta(x)=0$ for all $x$. From now on, the prefix codes we use are Huffman codes. Using their structure, we can efficiently implement our coding schemes.
An Huffman code will be characterized by a mapping $H_M^{\lambda}$, $H_M^{\lambda}:(1, \ldots ,M)\rightarrow (l_1, \ldots, l_M)$, such that $H_M^{\lambda}(i)= l_i, l_i\in \{1,\ldots,\log(\lambda)\}$ where $\lambda=2^l$ for some positive integer $l\leq M$. The $l_i$'s represents the lengths of codewords of some Huffman code with $M$ codewords and maximum codeword length of $\log(\lambda)$. We call $H_m^{\lambda}(i)$ the length function of the Huffman code. It is well known that $H_M^{\lambda}(i)$ is indeed a legitimate length function of some Huffman code if and only if $\sum_{i=1}^M 2^{-H_M^{\lambda}(i)}=1$. 
Notice that from our point of view, all Huffman codes with the same length function or equivalently, the same length set, have the same functionality, thus considered as the same code. Given some length set, it is of no importance, of course, which Huffman codebook will be actually used for encoding. Building an Huffman codebook from a length function is straightforward. We will use the scheme described in Theorem 2 for creating the sequential source code.
\subsubsection{Definition of the reference set of source codes}
We define $\mathcal{H}_M(\lambda)$ as the set of all Huffman codes (or equivalently, of all Huffman length sets) with $M$ codewords and maximal length of $\log(\lambda)$.
Our reference set of source codes is $\mathcal{H}_M(\lambda)$. Each encoder is defined by a mapping $e(i)=b_i,i\in\{1, 2, \ldots, M\}$. $\{b_1,b_2,\ldots,b_M\}$ is some Huffman codebook with length set $\{l(b_1),l(b_2),\ldots,l(b_M)\}$. As was explained, the actual codebook can be chosen arbitrarily among all codebooks which share the same length set. The corresponding decoder is, of course, defined by $d(b_i)=i,i\in\{1, 2, \ldots, M\}$.

\subsubsection{Graphical representation of all Huffman codes with maximal length $\log(\lambda)$}
\begin{figure}[hc]
	\centering 
		\includegraphics [width=4in, height=4in]{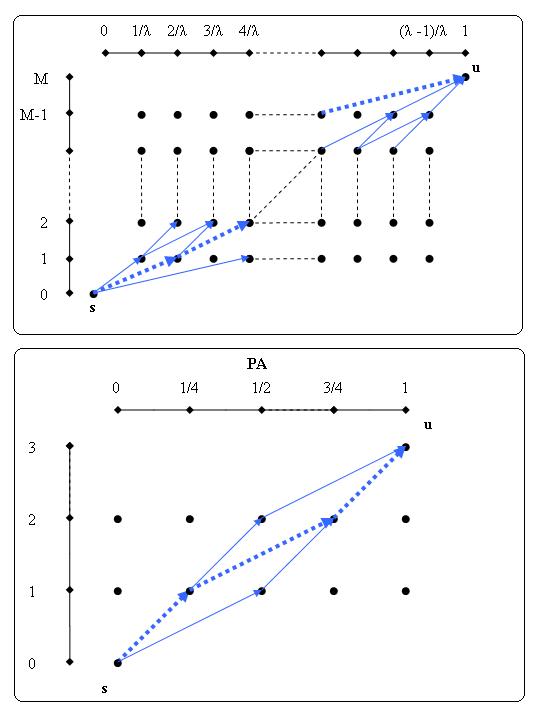} 
	\caption {The above figure describes a general graph. Below is a directed graph that represents $H_3(4)$, i.e., all Huffman codes with $M=3$ codewords and $\lambda=4$. The dashed path represents the probability function $\{\frac{1}{4},\frac{1}{2},\frac{1}{4}\}$ which is equivalent to the length set $\{l_1=2,l_2=1, l_3=2\}$. Remember that only edges with $(\hat{q}-q)$ equal to negative power of $2$ are legal.}
\label{huff_fig}
\end{figure}

Our next step is to reduce the problem of designing our source code (in other words, choosing randomly $H_k\in\mathcal{H}_M(\lambda)$, for each $k$, $k\in\{1, 2, \ldots, N\}$ given $x^{(k-1)l}$) to the problem of choosing randomly a path on an a-cyclic directed graph.\\ 
We describe each Huffman code as a path $r$ on a graph in the following way (see Fig. 2):\\ 
We use the following notation:\\
$\mathcal{V}$ - The set of all vertices:\\
$\{\frac{1}{\lambda}, \frac{2}{\lambda}, \ldots ,\frac{\lambda-1}{\lambda}\} \times \{1, 2, \ldots ,M-1\} \cup (0,0) \cup (1,M)$\\
$\mathcal{E}$ - The set of all edges:\\
$\{ ((q,j-1),(\hat{q},j)) : q,\hat{q}\in\{0,\frac{1}{\lambda},\frac{2}{\lambda},\ldots ,1\}, \hat{q}>q,\log\frac{1}{(\hat{q}-q)}\in \mathbb{Z}^+, j\in \{1,2, \ldots ,M\}\}$\\
$s$ - The starting point at the bottom left, i.e. $(0,0).$\\
$u$ - The end point at the top right, i.e. $(1,M)$.\\
$\mathcal{E}_z$ - The set of all edges starting from vertex $z$.\\
A general graph and the graph for all Huffman codes of order $3$ with $\lambda=4$, i.e., with maximal length $l=2$ are described in Fig.{ \ref{huff_fig}}.
The horizontal axis represents the Probability Axis (PA) $[0,1]$.
The vertical axis represents the $M-1$ choices needed for dividing the PA into $M$ segments.
A path composed of the edges $\{ (0,0), (q_1,1)\ldots , (q_{M-1},M-1),(1,M)\}$ represents $M-1$ consecutive choices of $M-1$ points $(q_1,\ldots,q_{M-1})$ which divide the PA into $M$ segments, creating discrete probability distribution, with $M$ probabilities.
Each edge on a path represents one choice, the choice of the next point on the PA, which defines the next segment. 
An edge $((q,j-1),(\hat{q},j))$ matches to the segment $[q,\hat{q})$ on the PA, thus equivalent to the probability $(\hat{q}-q)$.  
Each path is thus equivalent to some probability distribution. The correspondence between a probability distribution and a source code is as follows:
each probability distribution $\{p_1, p_2, \ldots, p_M\}$, corresponds to the binary prefix code which has the length set $\{\left\lceil \log(\frac{1}{p_1})\right\rceil,\left\lceil \log(\frac{1}{p_1})\right\rceil,\ldots,\left\lceil \log(\frac{1}{p_M})\right\rceil\}$, i.e., to its suitable Shannon code. 
Therefore, in order to represent all the Huffman codes and only them, we allow only partitions which divides the PA to negative integer powers of 2. This is implemented in the following way:\\
First, we consider only edges $( (q,m),(\hat{q},m+1))$ such that $\log\frac{1}{(\hat{q}-q)}\in \mathbb{Z}^+$ as mentioned above, meaning that we choose only symbol probabilities from the type $p_m=2^{-j}, j\in\mathbb{Z}^+,m\in\{1, 2, \ldots, M\}$.
We get $|\mathcal{E}|<M\lambda\log(\lambda)$ edges. It is easy to see that part of the edges are not members of at least one full path from $s$ to $u$, thus of no use. In order to get rid of the unnecessary edges, we simply start from the edges which end on $M-1$ vertically, and erase those with no edges starting from them. Then we move down to row $M-2$, and so on.
This process takes $O(M\lambda\log(\lambda))$ time and is done once and off-line. After ``cleaning" the graph, we have a graph that contains all the probability functions from the type 
$P: P(m)=2^{-j_m}, j_m\in\mathbb{Z}^+,m\in\{1, 2, \ldots, M\}$ and only them. It is well known that for probabilities of this type, the length set of the corresponding Huffman code is identical to that of the Shannon code and is simply $\{\log(\frac{1}{p_1}),\log(\frac{1}{p_2}),\ldots,\log(\frac{1}{p_M})\}$.
Each path matches uniquely to a specific Huffman code from the set $\mathcal{H}_M(\lambda)$ and $vice\ versa$, so the graph cover all Huffman codes with different length set and maximal codeword length of $\log(\lambda)$ and only them.
For each edge $a\in\mathcal{E}$ and time $t$, we assign a weight $\delta_{a,t}$:
\begin{equation}
\begin{array}{ll}
\delta_{a,t}=\exp\{-\eta f_t(j)\log\frac{1}{(\hat{q}-q)}\}=\exp\{-\eta f_t(j)l(q,\hat{q})\} ,&a=((q,j-1),(\hat{q},j))
\end{array}
\label{huff_gw}
\end{equation}
where $f_t(j)$ is the empirical relative frequency of the $j$th input symbol at time $t$, i.e., the number of times this symbol appears in the input sequence $x^t$. It can be seen from (\ref{huff_gw}) that a weight $\delta_{a,t}$ depends only on the horizontal coordinates of the edge $a$, thus we can denote it as $\delta_{(q,\hat{q}),t}$.
The cumulative weight of a path $r=\{ (0,0), (q_1,1)\ldots , (q_{M-1},M-1),(1,M)\}$ at time $t$ is defined as the product of its edges' weights:
\begin{equation}
\Lambda_{r,t}=\prod_{a\in r} \delta_{a,t}=\exp\{-\eta \sum_{a\in r} f_t(j)l(q,\hat{q})\}
\end{equation}
The sum is exactly the cumulative length of the encoding of the string $x^t$, using the Huffman code represented by this path, i.e., the length of $b_t$.
Again, following the WPA, we define:
\begin {equation}
G_t(z)=\sum_{r\in \mathcal{R}_z} \prod_{a\in r}\delta_{a,t}
\end{equation}
where $z$ is a vertex on the graph, $\mathcal{R}_z$ is the set of all paths from $z$ to $u$ and $a$ is an edge on the path $r$.
Continuing exactly as in Subsection 3.1.2, we efficiently implement the random choices of codes according to (\ref{enc_choice2}).\\ 
The procedure of updating the weights and finding the next code a total of $O(|\mathcal{E}|)+O(M)<O(M\lambda\log(\lambda))+O(M)$. Now, it is easy to see that it suffices to take $\lambda =\min{(2^{M-1},n)}$ to get all relevant Huffman codes of order $M$ when the sequence length is $n$. The procedure is repeated at the beginning of each data block, giving a total computational complexity of $O(n/l\cdot M\lambda\log(\lambda))+O(nM/l)$.
As a special case of the general LC scheme, we get the following result:
\begin{corollary}
Let $\mathcal{H}_M(\lambda)$ be the set of Huffman codes we defined above. Then there exists a sequential source code $(\tilde{e},\tilde{d})$ such that for all $x^n\in{\cal{X}}^n$:
\begin{equation}
\mathbb{E}\{\frac{1}{n}[{\cal{L}}^n_{(\tilde{e},\tilde{d})}({x^n})-\min_{(e',d')\in \mathcal{H}_M(\lambda)}{\cal{L}}^n_{(e',d')}(x^n)]\}\leq\\ M\sqrt{\log|\mathcal{H}_M(\lambda)|/2}(n/l)^{-\frac{1}{2}}
\end{equation}
Moreover, the scheme can be implemented with computational complexity of $O(n/l \cdot M\lambda\log(\lambda))+O(n/l\cdot M)$.
\end{corollary}
The proof is similar to that of the Theorem 2, with
\begin {equation}
{\cal{L}}(x_t)=l(b_t)\Rightarrow B=M-1,{\cal{L}}^n_{(e,d)}(x^n)=\sum_{t=1}^n l(b_t)
\end {equation}
and with one additional difference: suppose we use a randomization sequence $\{U_i\}$ of i.i.d. random variables, uniformly distributed on $[0,1]$, for implementing the random choices used in our WPA.
If we assume that the decoder also has access to this sequence, there is no need to inform it the identity of the encoder. Since the encoding is lossless, the decoder has all the information about the past. Therefore, given the randomization sequence, the decoder can achieve the identity of the next source code by itself.\\ 
Also notice that in this case,  in choosing $l$, there is a trade-off between convergence and the computational complexity. Choosing a small $l$ will improve the upper bound, but on the other hand, will increase the complexity.\\
In order to get some feeling about the compression performance of this scheme, we give the following example.\\
$\mathbf{Example.}$ : if we have $n=10^{10}$, $M=256$ and we take $l=\log(n)$, $\lambda=n$ so we have $\mathcal{H}_M(\lambda)<[\log(n)]^M$, we obtain that the difference between the best static Huffman code for $x^n$ and our scheme is less than $0.3$ bit per symbol.

\noindent\emph{Formal description of the on-line algorithm}: Using the set of encoders described above, we now have the following algorithm:
\begin{enumerate}
	\item Choose $l$, the length of a data block, and let $K = n / l$.
	\item Initialize $k$ to $0$.
	\item Build the encoders graph as described in this section.
	\item Initialize all the weights $\delta_{a,0}$ to 1.
	\item At the beginning of block no. $k$, i.e. at time $t_k=kl+1, k=\{1,\ldots,K\}$, calculate $\delta_{q,\hat{q},t_k}$ for each pair 	$(q,\hat{q})$ according to (\ref{huff_gw}).
	\item Update the weights of all edges to the new $\delta_{(q,\hat{q}),t_k}$'s.
	 \item Calculate $G_{t_k}(z)$ recursively, for all $z$, according to (\ref{g_t_recursive}).
	 \item Choose the encoder $e_k$ randomly as described in Subsection 3.1.2, using (\ref{seq_decis_prob}).
   \item Encode the next block, using the chosen expert $e_k$:\\
   $b_i= e_k(x_i),\ k•l\leq i \leq (k+1)•l-1$
   \item If $k<K$, increment $k$ and go to 5.     
\end{enumerate}
\subsection {An efficient adaptive LC scheme for the WZ case}
In this subsection, we return to the general LC scheme and the reference set of source codes defined in Subsection 4.2. We combine the WZ scheme from Subsection 3.1 and the Huffman coding from the previous subsection into one efficient LC scheme. One interesting special case of the following scheme, is described in appendix A. This special case is obtained by degenerating the side information alphabet into alphabet of size one. 
\subsubsection{Definition of the reference set of source codes}
The Input Alphabet Axis (IAA) is defined as the $|{\cal{X}}|$-sized vector $(1, 2, \ldots, |{\cal{X}}|)$. A division of the IAA is given by the $(M-1)$-sized increasing sequence $\mathbf{r}=(z_1,\ldots,z_{M-1}), z_i\in \{1,2,\ldots,|{\cal{X}}|-1\}$. $z_0$ and $z_M$ are defined to be $0$ and $|{\cal{X}}|$ respectively. Each combination between specific division $r$ and a specific Huffman length function defines a specific encoder in the following way:
\begin{equation}
\begin {array} {llll}
e(x)= b_i:z_{i-1} < Num(x) \leq z_i,&i\in \{1,\ldots,M\},&b_i\in \{b_i\}_{i=1}^{M},&\{l(b_i)\}_{i=1}^{M}\in {\cal{H}}_M(\lambda)
\end{array}
\end{equation}
We define $E\times {\cal{H}}_M(\lambda)$ as the set of all encoders which obtained by such combination.
\subsubsection{Graphical representation of the set of encoders}
The random choice of the encoders can be done efficiently using in this case, a Three-Dimensional (3D) a-cyclic directed graph instead of 2D.
We use the following notation:\\
$\mathcal{V}$ - The set of all vertices:\\
$\{1, 2,\ldots ,|{\cal{X}}|-1\}\times \{\frac{1}{\lambda}, \frac{2}{\lambda}, \ldots ,\frac{\lambda-1}{\lambda}\}\times \{1, 2, \ldots ,M-1\} \cup (0,0,0) \cup (|{\cal{X}}|,1,M)$\\
$\mathcal{E}$ - The set of all edges:
$\{ [ (z,q,j-1),(\hat{z},\hat{q},j)] : z,\hat{z}\in\{0,1,2,\ldots ,|{\cal{X}}|\}\\ q,\hat{q}\in\{0,\frac{1}{\lambda},\frac{2}{\lambda},\ldots ,1\}, j\in\{1,2,\ldots ,M\},\hat{z}>z, \hat{q}>q , \log\frac{1}{(\hat{q}-q)}\in \mathbb{Z}^+\}$\\
$s$ - The starting point in the bottom left, i.e. $(0,0,0)$\\
$u$ - The end point in the top right, i.e. $(|{\cal{X}}|,1,M)$\\
$\mathcal{E}_z$ - The set of all edges starting from vertex $z$.\\
\begin{figure}[hc]
	\centering		\includegraphics[width=3.5in, height=3.5in]{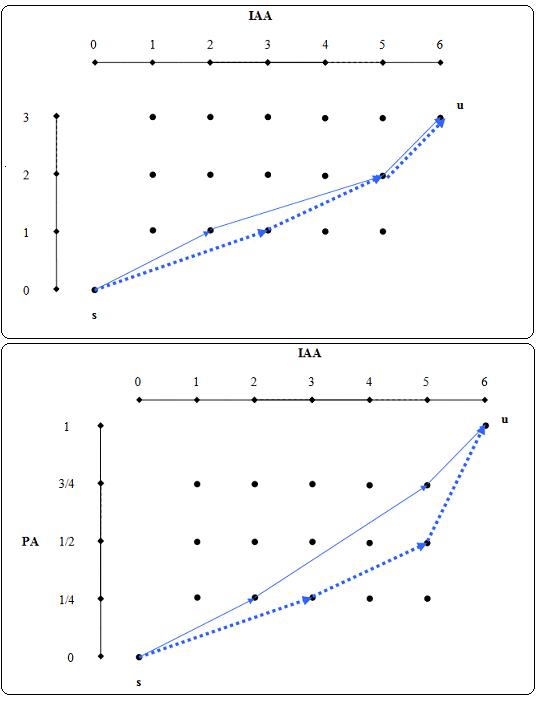} 
	\caption { {The lower graph presents the PA-IAA plane for $|{\cal{X}}|=6$, $M=3$ and $\lambda=4$. The dashed path represent the probability distribution with probabilities $\{\frac{1}{4},\frac{1}{4},\frac{1}{2}\}$ attached to the input alphabet subsets $\{\{1,2,3\},\{4,5\},\{6\}\}$ respectively. The other path represent the probability distribution $\{\frac{1}{4},\frac{1}{2},\frac{1}{4}\}$ attached to the partition $\{\{1,2\},\{3,4,5\},\{6\}\}$. The upper graph shows the same paths on the IAA - vertical axis plane. Remember that the vertical axis represent the $M-1$ consecutive decisions needed for defining an encoder.}}
\label{LC_fig}	
\end{figure}
The IAA represents the ordered input alphabet. The PA represents the probability axis $[0,1]$.
The vertical axis represents the $M-1$ choices needed for dividing the IAA and the PA simultaneously, each one into $M$ segments.
A path composed of the edges $\{ (0,0,0), (z_1,q_1,1)\ldots , (z_{M-1},q_{M-1},M-1),(|{\cal{X}}|,1,M)\}$ represents $M-1$ consecutive choices of $M-1$ points $([z_1,q_1],\ldots,[z_{M-1},q_{M-1}])$ on the PA-IAA plane, which divide the IAA and the PA into $M$ segments, creating $M$ subsets of the input alphabet, and $M$ probabilities.
Each edge on a path represents one choice, the choice of the next point on the horizontal subspace, which defines the next segment on the IAA and the next point on the PA. 
An edge $((z,q,j-1),(\hat{z},\hat{q},j))$ matches to the segment $(z,\hat{z}]$ on the IAA and to the segment $(q,\hat{q}]$ on the PA, thus equivalent to the subset $\{x:z < x \leq \hat{z}\}$ when assigned the probability $(\hat{q}-q)$. Therefore, a path from $s$ to $u$, having $M$ edges, defines partition of the input alphabet into $M$ subsets, each subset assigned a probability. Each probability $(\hat{q}-q)$ is equivalent to the length $\log(\frac{1}{\hat{q}-q})$, which is the length of a codeword in a real Huffman code as was explained in detail in Subsection 5.1. Therefore, each partition defines a specific encoder, which implements the alphabet division $\mathbf{r}=(z_1,\ldots,z_{M-1})$ and uses an Huffman code with the length set $\{\log(\frac{1}{q_1-q_0}),\ldots,\log(\frac{1}{q_M-q_{M-1}})\}$ for the variable rate coding part, where the lengths assigned to the subsets respectively.
As in the case with no distortion, we ``clear" the graph at off-line from edges with no use. This is done in the same way described in Subsection 5.1.2.
After that, we end up with $|\mathcal{E}|< M |{\cal{X}}|^2 \lambda\cdot \log(\lambda)$. An example of a PA-IAA plane is presented in Fig.{ \ref{LC_fig}}.
For each edge $a\in\mathcal{E}$ and time $t$, we assign a weight $\delta_{a,t}$:
\begin{equation}
\begin{array}{ll}
\delta_{a,t}=\delta_{(z,\hat{z}),t}\cdot\exp\{-\eta\cdot \delta\cdot f_t(z,\hat{z})l(q,\hat{q})\} ,&a=((z,q,j-1),(\hat{z},\hat{q},j))
\end{array}
\label{delta_lc}
\end{equation}
where $\delta_{(z,\hat{z}),t}$ is given by (\ref{lam_wei}), and $f_t(z,\hat{z})$ is given by:
\begin{equation}
f_t(z,\hat{z})=\sum_{i=1}^t I_{x_i\in (z,\hat{z}]}
\label{f_t_def}
\end{equation}
which is the empirical frequency of the subset $\{x:z < x \leq \hat{z}\}$ in the input sequence $x^t$. 
The cumulative weight of a path $\{ (0,0,0), (z_1,q_1,1)\ldots , (z_{M-1},q_{M-1},M-1),(|{\cal{X}}|,1,M)\}$ at time $t$ is the product of its edges' weights:
\begin{equation}
\Lambda_{r,t}=\prod_{a\in r} \delta_{a,t}
\end{equation}
$\Lambda_{r,t}$ is simply $F^{\cal{L}}_{e,t}$:
\begin{equation}
\begin{array}{lll}
\Lambda_{r,t}&=&\prod_{a\in r} \delta_{a,t}\\&=&\prod_{m=1}^M\{\delta_{(z,\hat{z}),t}\cdot\exp\{-\eta\cdot \delta\cdot f_t(z_{m-1},z_m)l(q_{m-1},q_m)\}\}\\&=&\prod_{m=1}^M \delta_{(z,\hat{z}),t}\prod_{m=1}^M\exp\{-\eta\cdot \delta\cdot f_t(z_{m-1},z_m)l(q_{m-1},q_m)\}\\&=&F_{e}\cdot\gamma_{e,t}
\end{array}
\end{equation}
where the last equality follows from (\ref{fe_is_product}) and the definition of $\gamma_{e,t}$.
Following the WPA exactly as we did in the previous parts, we implement efficiently the random choice of the encoder. The random choice of one of the possible decoders given the encoder, remains the same, as was shown before.

\noindent\emph{Formal description of the on-line algorithm}: Using the set of encoders described above, we now have the following algorithm:
\begin{enumerate}
	\item Calculate $l$, the optimal length of a data block, according to (\ref{l_eta_def_2}), and let $K = n / l$.
	\item Initialize $k$ to $0$, and all the weights $\lambda_{x,y,0}$ to $1$.
	\item Build the encoders graph as described in this section.
	\item Initialize all the weights $\delta_{a,0}$ to 1.
	\item At the beginning of block no. $k$, i.e. at time $t_k=kl+1$, update the weights in the following way:\\
	$\lambda_{x,y,t_k} = \lambda_{x,y,t_{k-1}}\exp(\eta  \sum_{i=(k-1)•l+1}^{k•l}I_{x_i=x}  P_{Y|X}(x,y) )$
	\item At the beginning of block no. $k$, calculate $\delta_{z,\hat{z},t_k}$ and $f_{t_k}(z,\hat{z})$ for each pair $(z,\hat{z})$ according to (\ref{lam_wei}) and (\ref{f_t_def}), respectively.
	\item Update the weights of all edges to the new $\delta_{a,t_k}$'s according to (\ref{delta_lc}).
	 \item Calculate $G_{t_k}(z)$ recursively, for all $z$, according to (\ref{g_t_recursive}).
	 \item Choose the encoder $e_k$ randomly as described in Subsection 3.1.2, using (\ref{seq_decis_prob}).
    \item For each pair $(z,y)$, choose the decoder function $d_k$ randomly according to (\ref{dec_calc}).
   \item Use the first $\left\lceil \log(N)\right\rceil$ bits at the beginning of the $k$th block to inform the decoder the identity of $d_k$, chosen in the previous step, where $N$ is the number of experts.
   \item Encode the next block using the chosen expert $e_k$:\\
   $b_i= e_k(x_i),\ k•l+\log(N)+1 \leq i \leq (k+1)•l-1$
   \item If $k<K$, increment $k$ and go to 3.  
\end{enumerate}
The total complexity of the algorithm is $O(n/l\cdot|{\cal{X}}|^3)+O(n/l\cdot M|{\cal{X}}|^2\lambda\cdot\log(\lambda))+O(n/l|{\cal{X}}|^2)+O(n)$.
\subsection{General distortion measures}
Let $\rho(x,\hat{x})$ be some bounded distortion measure.
Given an encoder $e$, we define:
\begin{equation}
\begin{array}{l}
\lambda_{x,y,e,t}=\exp\left\{-\eta\sum_{\{x':x'\neq x,e(x)=e(x')\}}n_t(x') P_{Y|X}(y|x')\rho(x,x')\right\}
\label{gen_lam}
\end{array}
\end{equation}
It is easy to see that given some possible decoder $d$, the distortion of the pair $r=(e,d)$ is:
\begin{equation}  
\lambda_{r,t}=\prod_{y\in{\cal{X}}}\prod_{z=1}^M\lambda_{d(z,y),y,e,t}
\end{equation}
using the generalized $\lambda$'s we defined, we can continue exactly as in the Hamming case. 
Each generalized $\lambda_{x,y,e,t}$ contains an exponent of a sum of $O(|{\cal{X}}|)$ products. So given an encoder the complexity is increased by a factor of $|{\cal{X}}|$. 
An example of using a general distortion measure is in the next subsection.
\subsection{Variable-Rate coding - Quantizers with Huffman codes}
In this subsection, we describe a special case of the variable-rate coding scheme of Section 5.2.
The ordered input alphabet is now composed of points on the real axis ${\cal{X}}=\{0,\frac{1}{K},\frac{2}{K},\ldots,1\}$ where $K>2$ is some positive integer. It is easy to see that $|{\cal{X}}|=K+1$. In this part, we assume there is no side information or equivalently, that the side information alphabet is of size one.
As described in 5.2, each encoder partitions the input alphabet into $M$ subsets and use some Huffman code for the lossless coding part.
We use some general bounded distortion measure which satisfies:
\begin {equation}
\rho(x,\hat{x})=\rho(|x-\hat{x}|)
\end {equation}
Under these conditions, our set is actually a set of quantizers of size $M$, where the points of each quantizer are encoded with some Huffman code.
A source code $(e,d)$ is called a Nearest-Neighbor (NN) quantizer if for all $x$ it satisfies:
\begin {equation}
\rho(|x-d(e(x))|)=\min_{x'\in {\cal{X}}} \rho(|x-x'|)
\end {equation}
By definition, the distortion of a NN quantizer is always the minimal among all quantizers with the same points.
It is easy to see that all the possible NN quantizers for this case, are included in our reference set of source codes. 
The cumulative LC is:
\begin {equation}
{\cal{L}}^n_{(e,d)}(x^n)=\sum_{t=1}^n {\cal{L}}(x_t)= \sum_{t=1}^n \rho(x_t,d(e(x_t)))+\delta\sum_{t=1}^n l(e(x_t))
\end {equation}
We call our set of encoders $Q\times{\cal{H}}$. We build an a-cyclic directed graph according to the description in Subsection 5.2. For each edge $a\in\mathcal{E}$ and time $t$, we assign a weight $\delta_{a,t}$, where in $\delta_{(z,\hat{z}),t}$ we substitute the generalized $\lambda$'s defined in (\ref{gen_lam}):
\begin{equation}
\lambda_{x,(z,\hat{z}],t}=\exp\left\{-\eta\sum_{\{x':x'\neq x, x'\in(z,\hat{z}]\}}n_t(x')\rho(x,x')\right\}
\end{equation}
Notice that the dependency on $y$ was omitted, and that the dependency on $e$ was replaced by $(z,\hat{z}]$. This stems from the fact that the subset $\{x':e(x)=e(x')\}$ in (\ref{gen_lam}), is equal to the subset $\{x': x'\in(z,\hat{z}]\}$ in our case, by the definition of our graph.
After choosing an encoder, choosing a decoder is done according to (\ref{dec_calc}) where again, we use the generalized $\lambda$'s, and the dependency on $y$ is omitted. The complexity of the algorithm remains the same as in Subsection 5.2.

\end{document}